

\input harvmac

\overfullrule=0pt


\def\H{{\scriptscriptstyle H}}
\def\I{{\scriptscriptstyle I}}
\def\J{{\scriptscriptstyle J}}

\def\L{{\scriptscriptstyle L}}

\def\P{{\scriptscriptstyle P}}
\def\Q{{\scriptscriptstyle Q}}
\def\R{{\scriptscriptstyle R}}
\def\S{{\scriptscriptstyle S}}
\def\T{{\scriptscriptstyle T}}


\def\CA{{\cal A}}

\def\CD{{\cal D}}

\def\CL{{\cal L}}

\def\CQ{{\cal Q}}


\def\e{\epsilon}
\def\g{\gamma}
\def\l{\lambda}

\def\o{\sigma}

\def\u{\mu}
\def\v{\nu}


\def\aEM{\alpha_{\scriptscriptstyle EM}}

\def\bar#1{\overline{#1}}
\def\ccdot{\hbox{\kern-.1em$\cdot$\kern-.1em}}
\def\cH{c_\H}
\def\cS{c_\S}
\def\cST{c_{\S\T}}
\def\dl{\overleftarrow{\partial}}

\def\dr{\overrightarrow{\partial}}

\def\EM{{\scriptscriptstyle EM}}

\def\gfive{\gamma^5}
\def\gtap{\raise.3ex\hbox{$>$\kern-.75em\lower1ex\hbox{$\sim$}}}
\def\HHCPT{{\scriptscriptstyle HHCPT}}
\def\HQET{{\scriptscriptstyle HQET}}
\def\hv{h_v^{(\Q)}}
\def\hvbar{\overline{h}_v^{(\Q)}}

\def\ltap{\raise.3ex\hbox{$<$\kern-.75em\lower1ex\hbox{$\sim$}}}
\def\LX{{\Lambda_\chi}}
\def\mb{m_b}

\def\mc{m_c}

\def\MeV{\> {\rm MeV}}
\def\mpi{m_{\pi}}

\def\mQ{m_\Q}
\def\muep{{\mu^{\epsilon/2}}}

\def\parenstar{{\scriptscriptstyle (*)}}
\def\proj{{1 + \slash{v} \over 2}}

\def\QCD{{\scriptscriptstyle QCD}}
\def\slash#1{#1\hskip-0.5em /}
\def\space{\>\>}
\def\Tsix{T_{\scriptscriptstyle (6)}}
\def\Tthree{T^{\scriptscriptstyle ({\bar 3})}}


\def\half{{1 \over 2}}

\def\sixth{{ 1\over 6}}
\def\third{{1 \over 3}}
\def\threehalves{{3 \over 2}}

\def\twothirds{{2 \over 3}}


\newdimen\pmboffset
\pmboffset 0.022em
\def\oldpmb#1{\setbox0=\hbox{#1}%
 \copy0\kern-\wd0
 \kern\pmboffset\raise 1.732\pmboffset\copy0\kern-\wd0
 \kern\pmboffset\box0}
\def\pmb#1{\mathchoice{\oldpmb{$\displaystyle#1$}}{\oldpmb{$\textstyle#1$}}
      {\oldpmb{$\scriptstyle#1$}}{\oldpmb{$\scriptscriptstyle#1$}}}
\def\pib{{\pmb{\pi}}}


\def\LongTitle#1#2#3#4{\nopagenumbers\abstractfont
\hsize=\hstitle\rightline{#1}
\vskip 1in\centerline{\titlefont #2}\centerline{\titlefont #3}
\centerline{\titlefont #4}
\abstractfont\vskip .5in\pageno=0}
%
%
\def\appendix#1#2{\global\meqno=1\global\subsecno=0\xdef\secsym{\hbox{#1.}}
\bigbreak\bigskip\noindent{\bf Appendix. #2}\message{(#1. #2)}
\writetoca{Appendix {#1.} {#2}}\par\nobreak\medskip\nobreak}
%

\nref\Wise{M. Wise, Phys. Rev. {\bf D45} (1992) R2188.}
\nref\Burdman{G. Burdman and J. Donoghue, Phys. Lett. {\bf B280} (1992) 287.}
\nref\Yan{T.M. Yan, H.Y. Cheng, C.Y. Cheung, G.L. Lin, Y.C. Lin and H.L.
  Yu, Phys. Rev. {\bf D46} (1992) 1148.}
\nref\ChoI{P. Cho, Phys. Lett. {\bf B285} (1992) 145.}
\nref\ChoII{P. Cho. HUTP-92/A039 (1992).}
\nref\GeorgiI{H. Georgi, Heavy Quark Effective Field Theory, {\it in} Proc.
  of the Theoretical Advanced Study Institute 1991, ed. R.K. Ellis, C.T. Hill
  and J.D. Lykken (World Scientific, Singapore, 1992) p. 589.}
\nref\GeorgiII{H. Georgi, Nucl. Phys. {\bf B348} (1991) 293.}
\nref\GeorgiIII{H. Georgi, Phys. Lett. {\bf B240} (1990) 447.}
\nref\FGL{A. Falk, B. Grinstein and M. Luke,
Nucl. Phys. {\bf B357} (1991) 185.}
\nref\GeorgiIV{
  H. Georgi, HUTP-92/A036 (1992)\semi
  A. Manohar and H. Georgi, Nucl. Phys. {\bf B234} (1984) 189\semi
  H. Georgi and L. Randall, Nucl. Phys. {\bf B276} (1986) 241.}
\nref\CLEO{CLEO Collaboration, F. Butler {\it et al.} Cornell Preprint CLNS
  92/1143.}
\nref\PDB{Review of Particle Properties, Phys. Rev. {\bf D45}, Part 2 (1992).}
\nref\IsgurWise{N. Isgur and M. Wise, Phys. Rev. {\bf D41} (1990) 151.}
\nref\FalkLuke{A. Falk and M. Luke, UCSD/PTH 92-14 (1992).}
\nref\Amundson{J.F. Amundson, C.G. Boyd, E. Jenkins, M. Luke, A.V. Manohar,
  J.L. Rosner, M.J. Savage and M.B. Wise, Caltech preprint CALT-68-1816
  (1992).}
%

\LongTitle{HUTP-92/A043}
  {Electromagnetic Interactions in}{Heavy Hadron Chiral Theory}{}
\centerline{Peter Cho
  \footnote{$^\dagger$}{Address after Sept 21, 1992: California Institute
  of Technology, Pasadena, CA  91125.} and Howard Georgi}
\bigskip\centerline{Lyman Laboratory of Physics}
\centerline{Harvard University}
\centerline{Cambridge, MA 02138}

\vskip .3in


      Electromagnetic interactions are incorporated into Heavy Hadron
Chiral Perturbation Theory.  Short and long distance magnetic moment
contributions to the chiral Lagrangian are identified, and $M1$
radiative decays of heavy vector mesons and sextet
baryons are studied.  Using recent CLEO $D^*$ branching fraction ratio data,
we fit the meson coupling to the axial vector Goldstone current and find
$g_1^2 = 0.34 \pm 0.48 $ for $\mc = 1700 \MeV$.  Finally, we obtain model
independent predictions for total and partial widths of charm and
bottom vector mesons.

\Date{9/92}

      A synthesis of Chiral Perturbation Theory and the Heavy Quark Effective
Theory (HQET) has recently been developed \refs{\Wise{--}\ChoII}.  This hybrid
effective theory describes low energy strong interactions between light
Goldstone bosons and hadrons containing a single heavy quark.  Weak $b \to c$
transitions among heavy meson or baryon states can also be incorporated
into this framework.  In this letter, we extend the theory's formalism to
include electromagnetism and then study the radiative decays of heavy vector
mesons and sextet baryons.

      To begin, we briefly review the basic elements of Heavy Hadron
Chiral Perturbation Theory (HHCPT).
\foot{This introductory discussion closely follows that presented
in refs.~\refs{\ChoI,\ChoII} to which we refer interested readers for
further details.}
The Goldstone bosons resulting from the chiral symmetry breakdown
$SU(3)_\L \times SU(3)_\R \to SU(3)_{\L+\R}$ appear in the pion octet
\eqn\pionoctet{\pib = \sum_{a=1}^8 \pi^a T^a = {1 \over \sqrt{2}}
\pmatrix{ \sqrt{\half} \pi^0 + \sqrt{\sixth} \eta & \pi^+ & K^+ \cr
\pi^- & - \sqrt{\half} \pi^0+\sqrt{\sixth}\eta & K^0 \cr
K^- & \bar{K}^0 & - \sqrt{\twothirds}\eta \cr}}
and are associated with the pion decay constant $f \approx 93 \MeV$.
These fields are arranged into the exponentiated matrix functions
$\Sigma=e^{2i \pib/f}$ and $\xi=\hbox{``}\sqrt{\Sigma}\hbox{''}
=e^{i \pib/f}$
that transform under the chiral symmetry group as
\eqn\Sigmafields{\eqalign{\Sigma &\to L \Sigma R^\dagger \cr
\xi &\to L \xi U^\dagger = U \xi R^\dagger. \cr}}
Here $L$ and $R$ represent global elements of $SU(3)_\L$ and $SU(3)_\R$, while
$U$ acts like a local $SU(3)_{\L+\R}$ transformation.
Chiral invariant terms that describe Goldstone boson self interactions
are constructed from the fields in \Sigmafields\ and their derivatives.

      Hadrons containing a heavy quark emit and absorb light
Goldstone bosons with no appreciable change in their four velocities.  They
are consequently described by velocity dependent fields.  In the meson sector,
we introduce the operators $P_i(v)$ and $P^*_{i\u}(v)$ that annihilate
pseudoscalar and vector mesons with quark content $Q \bar{q}$.  When the
suppressed heavy quark label carried by these fields corresponds to charm,
their individual components are given by
\eqn\Pops{\eqalign{
(P_1,P_2,P_3) & = (D^0,D^+,D^+_s) \cr
(P_1^*,P_2^*,P_3^*) &= ({D^*}^0,{D^*}^+,{D^*}^+_s).\cr}}
In the infinite quark mass limit, it is useful to combine the degenerate
meson spin states into the $4\times 4$ matrix field \refs{\Wise,\GeorgiI}
\eqna\Hfield
$$ \eqalignno{
H_i(v) &= \proj \bigl[ - P_i(v) \gfive + P^*_{i \u}(v) \gamma^\u \bigr] &
  \Hfield a \cr} $$
and its conjugate
$$ \eqalignno{\bar{H}^i(v) &= \gamma_0 H^\dagger \gamma_0 =
  \bigl[ {P^\dagger}^i(v) \gfive + {P^{*\dagger}_\u}^i(v) \g^\u \bigr] \proj.
  & \Hfield b\cr } $$
$H$ carries a heavy quark spinor index and a separate light antiquark
spinor index and transforms as an antitriplet under $SU(3)_{\L+\R}$ and
doublet under $SU(2)_v$.

      Baryons with quark content $Qqq$ enter into the theory in two types
depending upon the angular momentum of their light degrees of freedom.
In the first case, the light spectators are arranged in a symmetric spin-$1$
configuration that couples with the heavy spin-$\half$ quark to form
$J^P=\half^+$ and $J^P=\threehalves^+$ states.  When the heavy partner is
taken to be charm, the spin-$\half$ states are destroyed by the
Dirac operators appearing in the symmetric sextet representation
\eqn\Bops{{\bf S} = \sum_{\I=1}^6 S^\I \Tsix^\I =
\pmatrix{\Sigma^{++}_c & {\sqrt\half} \Sigma^+_c & {\sqrt\half} {\Xi^+_c}' \cr
{\sqrt\half} \Sigma^+_c & \Sigma^0_c & {\sqrt\half} {\Xi^0_c}' \cr
{\sqrt\half} {\Xi^+_c}' & {\sqrt\half} {\Xi^0_c}' & \Omega^0_c \cr} . }
Their spin-$\threehalves$ counterparts are annihilated by the corresponding
Rarita-Schwinger field $\bf S^*_\u$.  We again combine the Dirac and
Rarita-Schwinger operators into the ``super'' fields \GeorgiII\
\eqn\Sfield{\eqalign{
S^{ij}_\u(v) &= \sqrt{\third} (\g_\u+v_\u) \gfive S^{ij}(v)
  + {S^*_\u}^{ij}(v) \cr
\bar{S}_{ij}^\mu(v) &= - \sqrt{\third} \bar{S}_{ij}(v) \gfive
 (\g^\u+v^\u)+ {\bar{S}^{\, *}_{ij}}^\u(v)  . \cr}}
Then $S_\u$ transforms as a sextet under $SU(3)_{\L+\R}$, doublet under
$SU(2)_v$, and is an axial vector.

	The spectators in the second case are bound together into an
antisymmetric spin-$0$ state.  The resulting $J^P=\half^+$ baryons are assigned
to the field $T_i(v)$, which is an $SU(3)_{\L+\R}$ antitriplet and $SU(2)_v$
doublet.  When $Q=c$, the components of $T_i$ are the singly
charmed baryons
\eqn\Tfield{(T_1,T_2,T_3) = (\Xi_c^0, - \Xi_c^+, \Lambda_c^+).}
These antitriplet baryons can alternatively be arranged into the antisymmetric
matrix
\eqn\Tops{ {\bf T} = \sum_{i=1}^3 T_i \Tthree_i =
  \pmatrix{0 & {\sqrt\half} \Lambda^+_c & {\sqrt\half} \Xi^+_c \cr
  - {\sqrt\half} \Lambda^+_c & 0 & {\sqrt\half} \Xi^0_c \cr
  - {\sqrt\half} \Xi^+_c & -{\sqrt\half} \Xi^0_c & 0 \cr}}
where $\bigl( T_i^{({\bar 3})} \bigr)_{jk} = \e_{ijk}/\sqrt{2}$.

      We can now construct the zeroth order effective chiral Lagrangian that
describes the low energy interactions between
light Goldstone bosons and heavy hadrons in the infinite heavy quark
mass limit.  The leading order terms must be hermitian, Lorentz invariant,
light flavor and heavy quark spin symmetric, and parity even.  We can also
readily incorporate electromagnetism into the hybrid
chiral theory by gauging a $U(1)_\EM$ subgroup of the global
$SU(3)_\L \times SU(3)_\R$ symmetry group.  Only long wavelength
photons with energies less than the chiral symmetry breaking scale
explicitly remain in the low energy theory while short wavelength modes are
integrated out.  In $d=4-\e$ dimensions, the effective Lagrangian
looks like
\foot{Meson contributions are written in terms of the dimension-$\threehalves$
field $H'=\sqrt{M_\H} H$ so that all heavy mass dependence is removed from
the leading order Lagrangian.}
\eqna\Lzero
$$ \eqalignno{\CL^{(0)} &= - {1\over 4} F^{\u\v} F_{\u\v} +
  {\u^{-\e} f^2 \over 4} \Tr(\CD^\u \Sigma^\dagger
\CD_\u \Sigma) & \Lzero a \cr
\CL_v^{(0)} &= \sum_{Q=c,b} \Bigl\{
  -i\Tr \bigl( {\bar{H}'}^i v \cdot \CD H_i' \bigr)
  -i \bar{S}^\u_{ij} v\cdot\CD S^{ij}_\u + (M_\S - M_\T) \bar{S}^\u_{ij}
   S^{ij}_\u
  +i \bar{T}^i v \cdot \CD \, T_i & \cr
& \qquad\qquad + g_1 \Tr \bigl( H_i' (\slash{ A})^i_j \gfive
  {\bar{H}'}^j \bigr) + i g_2 \varepsilon_{\u\v\o\l} \bar{S}^\u_{ik}
  v^\v (A^\o)^i_j (S^\l)^{jk} & \cr
& \qquad\qquad + g_3 \Bigl[ \e_{ijk} \bar{T}^i (A^\u)^j_l S^{kl}_\u
  + \e^{ijk} \bar{S}^\u_{kl} (A_\u)^l_j T_i \Bigr] \Bigr\}. & \Lzero b \cr}$$
The Goldstone bosons explicitly couple to the matter fields through the
axial vector combination
\eqna\pioncurrs
$$ \eqalignno{{\bf A}^\u &= {i \over 2}(\xi^\dagger \CD^\u \xi
  - \xi \CD^\u \xi^\dagger). & \pioncurrs a\cr} $$
They also communicate via the vector field
$$ \eqalignno{{\bf V}^\u &= \half (\xi^\dagger \CD^\u \xi
  + \xi \CD^\u \xi^\dagger) & \pioncurrs b \cr} $$
that appears inside the heavy hadron covariant derivatives
\foot{We distinguish the photon field $\CA^\u$ from the axial vector Goldstone
current $\bf A^\u$ by writing the former in calligraphy type.  Similarly, we
let $\CQ$ represent the electric charge operator, which is different from
the heavy quark symbol $Q$.}
\eqn\heavyderivs{\eqalign{
\CD^\u H'_i &= \partial^\u H'_i - H'_j (V^\u)^j_i - i \muep e \CA^\u
  \bigl[ \CQ_Q H'_i-H'_j \CQ^j_i \bigr] \cr
\CD^\u S^{ij}_\v &= \partial^\u S^{ij}_\v + (V^\u)^i_k S^{kj}_\v
  +(V^\u)^j_k S^{ik}_\v -i \muep e \CA^\u
  \bigl[ \CQ_Q S^{ij}_\v + \CQ^i_k S^{kj}_\v + \CQ^j_k S^{ik}_\v \bigr] \cr
\CD^\u T_i &= \partial^\u T_i - T_j (V^\u)^j_i -i \muep e \CA^\u
  \bigl[ \CQ_Q T_i - T_j \CQ^j_i \bigr] . \cr}}
The remaining Goldstone covariant derivatives in \Lzero{a}\ and
\pioncurrs{}\ are given by
\eqn\Goldderivs{\eqalign{
\CD^\u \Sigma^i_j &= \partial^\u \Sigma^i_j - i \muep e \CA^\u
  \bigl[ \CQ^i_k \Sigma^k_j - \Sigma^i_k \CQ^k_j \bigr] \cr
\CD^\u \xi^i_j &= \partial^\u \xi^i_j - i \muep e \CA^\u
  \bigl[ \CQ^i_k \xi^k_j - \xi^i_k \CQ^k_j \bigr] \cr}}
where
\eqn\Qcharge{{\pmb\CQ} = \pmatrix{\CQ_1 && \cr & \CQ_2 & \cr && \CQ_3 \cr}
       = \pmatrix{\twothirds && \cr & -\third & \cr && - \third\cr}}
denotes the light quark electric charge matrix.

	Spin symmetry violating contributions to the chiral Lagrangian
enter at $O(1/\mQ)$.  Among these are heavy quark magnetic moment terms
which mediate $M1$ radiative transitions.  As we will see, these terms are
completely fixed by heavy quark number conservation.  This simple but
crucial observation allows one to use experimental meson decay
information to determine the parameter $g_1$.

	Recall that the photon gauge field couples to
the conserved current that counts heavy quark number in the underlying QCD
theory as well as in the low energy HQET and HHCPT.  The original QCD current
appears in its well-known Gordon decomposed form as
\eqn\QCDcurr{J^\QCD_\u = \bar{Q}(p') \gamma_\u Q(p) =
  {1 \over 2 \mQ} \bar{Q}(p') \bigl[ (p'+p)_\u + i \sigma_{\u\v} (p'-p)^\v
  \bigr] Q(p).}
Running down in energy to the heavy quark thresholds and invoking the
velocity superselection rule \GeorgiIII\ to set
$p^{(\prime)} =\mQ v + k^{(\prime)}$, one can match this tree level current
onto the corresponding HQET expression
\eqn\HQETcurr{J^\HQET_\u
  = \hvbar \bigl[ v_\u + {i \over 2 \mQ} (\dr_\u - \dl_\u)
  + {1 \over 2 \mQ} \sigma_{\u\v} (\dr\null^\v + \dl\null^\v) \bigr] \hv .}
The $\CA^\u$ gauge field couples to this current in the Lagrangian
\eqn\HQETLagrangian{
  \CL^{(\HQET)}_v = \sum_{\Q=c,b} \Bigl\{\hvbar (iv \ccdot D) \hv + a_1 O_1
  + a_2 O_2 + a_3 O_3 \Bigr\} }
where the $O(1/\mQ)$ $O_i$ operators are constructed from either symmetric
or antisymmetric combinations of two HQET covariant derivatives
$ D^\u = \partial^\u - i \muep g G_a^\u T_a - i \muep e \CA^\u \CQ$:
\eqna\Oops
$$ \eqalignno{
O_1 &= {1 \over 2\mQ} \hvbar (iD)^2 \hv & \Oops a \cr
O_2 &= {\muep g \over 4\mQ} \hvbar \sigma_{\mu\nu} T_a \hv G^{\mu\nu}_a
    & \Oops b \cr
O_3 &= {\muep e \CQ_\Q \over 4\mQ} \hvbar \sigma_{\mu\nu} \hv F^{\mu\nu}.
    & \Oops c \cr} $$
The $a_i$ coefficients of these dimension-five operators are thus fixed by
\HQETcurr\ and equal unity at lowest order \FGL.

      Running down further in energy from the heavy quark thresholds to the
chiral symmetry breaking scale $\LX$, we match the electromagnetic pieces of
operators $O_1$ and $O_3$ onto the following short distance contributions to
the HHCPT Lagrangian:
\eqn\shortdist{\eqalign{
  \CL^{(short)}_v &= \sum_{Q=c,b} \Bigl\{
  -{1\over 2\mQ}\Tr (\bar{H}'(i\CD)^2 H')
  - {\muep\CQ_\Q e(\mQ)\over 4\mQ}\Tr(\bar{H}'\sigma_{\u\v} H' F^{\u\v} )\cr
& \qquad\qquad - {1 \over 2 \mQ} \bar{S}^\l_{ij} (i\CD)^2 S^{ij}_\l
  - {\muep\CQ_\Q e(\mQ)\over 4\mQ}
   \bar{S}^\l_{ij}\sigma_{\u\v}S^{ij}_\l F^{\u\v} \cr
& \qquad\qquad + {1 \over 2 \mQ} \bar{T}^i (i\CD)^2 T_i
  + {\muep \CQ_\Q e(\mQ) \over 4 \mQ} \bar{T}^i \sigma_{\u\v} T_i F^{\u\v}
  \Bigr\}. }}
These $O(1/\mQ)$ terms describe the interaction of photons with
the heavy quark constituent inside a $H'$, $S$ or $T$ hadron.  Consequently,
the Lorentz and flavor indices for the hadrons' light degrees of
freedom are trivially contracted.  Heavy quark number conservation
determines the ratio of the operator coefficients in \shortdist\
to the kinetic terms in the zeroth order Lagrangian \Lzero{b}. The
conserved HQET current therefore matches onto
\eqn\HHCPTcurr{\eqalign{J^\HHCPT_\u =
& -\Tr {\bar{H}'}^i \Bigl[ v_\u + {i \over 2 \mQ} (\dr_\u - \dl_\u)
 + {1 \over 2 \mQ} \sigma_{\u\v} (\dr\null^\v + \dl\null^\v) \Bigr] H'_i \cr
& - \bar{S}^\l_{ij} \Bigl[ v_\u + {i \over 2 \mQ} (\dr_\u - \dl_\u)
 + {1 \over 2 \mQ} \sigma_{\u\v} (\dr\null^\v + \dl\null^\v) \Bigr] S_\l^{ij}
  \cr
& + \bar{T}^i \Bigl[ v_\u + {i \over 2 \mQ} (\dr_\u - \dl_\u)
 + {1 \over 2 \mQ} \sigma_{\u\v} (\dr\null^\v + \dl\null^\v) \Bigr] T_i \cr}}
in the low energy chiral theory.

      Photons also couple to the light brown muck inside heavy
hadrons leaving the spins of their heavy quark constituents unaltered.
Such long distance interactions generate additional electromagnetic
contributions to the effective Lagrangian at the $\LX$ scale.  We focus
upon just the induced magnetic moment terms:
\eqn\longdist{\eqalign{
  \CL_v^{(long)} = {\muep e(\LX) \over \LX} & \Bigl\{
  \cH \Tr \bigl( {\bar{H}'}^i H'_j (-\CQ)^j_i \sigma_{\u\v} F^{\u\v} \bigr)
  + i \cS \bar{S}_{\u,ij} \bigl( \CQ^i_k S^{kj}_\v +
  \CQ^j_k S^{ik}_\v \bigr) F^{\u\v} \cr
  & + \cST \Bigl[ \e_{ijk} \bar{T}^i v_\u Q^j_l S^{kl}_\v +
  \e^{ijk} \bar{S}_{\v,kl} v_\u Q^l_j T_i \Bigr] F^{\u\v} \Bigr\}.}}
A few points about these long distance operators should be noted.  Firstly,
the suppressed heavy quark spinor indices in these spin symmetry preserving
terms are simply contracted.  Their light Lorentz and flavor indices on the
other hand are nontrivially arranged.  Secondly, the coefficients $\cH$,
$\cS$ and $\cST$ are {\it a priori} unknown.  But naive dimensional analysis
suggests that they are of order one \GeorgiIV. Finally, there is no long
distance magnetic moment interaction for just the antitriplet baryon since
the photon field cannot couple to its spinless light degree of freedom.

      Having identified the short and long distance magnetic moment
terms in the low energy chiral theory, we can now study $M1$ radiative
transitions between meson and baryon states.  Since the
hyperfine splitting between charmed pseudoscalar and vector meson
partners is only slightly greater than a pion mass, the electromagnetic
decay $D^* \to D \gamma$ competes with the strong process $D^* \to D \pi$.
The greater phase space for the electromagnetic transition offsets its
inherently smaller amplitude.  Bottom vector mesons must radiatively decay
because pion emission is kinematically forbidden.  So these $M1$ meson
processes are of genuine phenomenological interest.  Similar considerations
apply to the baryon transitions.

      The vector meson and sextet baryon radiative decay rates are readily
determined from the magnetic moment terms in \shortdist\ and \longdist:
\eqna\EMdecayrate
$$ \eqalignno{
&\Gamma(P^*_i \to P_i \gamma) = & \EMdecayrate a \cr
&\qquad \twothirds \Bigl({M_\P \over M_{\P^*}}\Bigr)
  \Bigl({M_{\P^*}^2 - M_\P^2 \over M_{\P^*}} \Bigr)^3
  \Bigl[ {\CQ_\Q\over 4\mQ} \aEM(\mQ)^{1/2} +
    {\cH\over\LX} \CQ_i \aEM(\LX)^{1/2} \Bigr]^2 & \cr
&\Gamma({S^*}^\I \to S^\I \gamma) =& \EMdecayrate b \cr
&\qquad {1 \over 18}
  \Bigl({M_\S \over M_{\S^*}}\Bigr)
  \Bigl({M_{\S^*}^2 - M_\S^2 \over M_{\S^*}} \Bigr)^3
  \Bigl[ {\CQ_\Q \over\mQ} \aEM(\mQ)^{1/2} +
  2{\cS\over\LX} \Tr({\Tsix^\I}^\dagger {\pmb \CQ} \Tsix^\I) \aEM(\LX)^{1/2}
  \Bigr]^2 & \cr
&\Gamma({S^\parenstar}^\I \to T_j \gamma) = &\EMdecayrate c \cr
&\qquad {1 \over 6}
  \Bigl({M_\T \over M_{\S^\parenstar}} \Bigr)
  \Bigl({M_{\S^\parenstar}^2 - M_\T^2 \over M_{\S^\parenstar}} \Bigr)^3
  \Bigl[ {\cST\over\LX} \Tr({\Tthree_j}^\dagger {\pmb \CQ} \Tsix^\I)
  \aEM(\LX)^{1/2} \Bigr]^2. & \cr} $$
One can clearly identify the short and long distance contributions to these
partial widths from their electric charges and associated inverse mass scales.
The corresponding strong interaction decay rates are derived from the Goldstone
axial vector couplings in the leading order Lagrangian \Lzero{b}:
\eqna\strongdecayrate
$$ \eqalignno{&\Gamma(P_i^* \to P_j \pi^a) = & \strongdecayrate a \cr
&\qquad {g_1^2 \over 48\pi f^2}
  \Bigl({M_\P \over M_{\P^*}}\Bigr)
  \Biggl[ {[M_{\P^*}^2-(M_\P+\mpi)^2][M_{\P^*}^2-(M_\P-\mpi)^2]
   \over M_{\P^*}^2} \Biggr]^{3/2} | (T^a)_{ji} |^2 & \cr
&\Gamma({S^*}^\I \to S^\J \pi^a) = & \strongdecayrate b \cr
&\qquad {g_2^2 \over 144\pi f^2}
  \Bigl({M_\S \over M_{\S^*}} \Bigr) \Biggl[
  {[M_{\S^*}^2-(M_\S+\mpi)^2][M_{\S^*}^2-(M_\S-\mpi)^2]
   \over M_{S^*}^2} \Biggr]^{3/2}
  |\Tr({\Tsix^\J}^\dagger T^a \Tsix^\I)|^2 & \cr
&\Gamma({S^\parenstar}^\I \to T_j  \pi^a) = & \strongdecayrate c \cr
&\qquad {g_3^2 \over 24\pi f^2}
  \Bigl({M_\T \over M_{\S^\parenstar}} \Bigr) \Biggl[
  {[M_{\S^\parenstar}^2-(M_\T+\mpi)^2][M_{\S^\parenstar}^2-(M_\T-\mpi)^2]
   \over M_{S^\parenstar}^2} \Biggr]^{3/2}
  |\Tr({\Tthree_j}^\dagger T^a \Tsix^\I)|^2. & \cr} $$

      None of the heavy hadron electromagnetic and strong partial widths
have been directly measured.  However, values for $D^*$ branching
fraction ratios are known \CLEO:
\eqn\branchingfractionratios{\eqalign{
R_\gamma^0 &= {\Gamma({D^*}^0 \to D^0 \gamma) \over \Gamma({D^*}^0 \to
  D^0 \pi^0)} = 0.572 \pm 0.057 \pm 0.081 \cr
R_\gamma^+ &= {\Gamma({D^*}^+ \to D^+ \gamma) \over \Gamma({D^*}^+ \to
  D^+ \pi^0)} = 0.035 \pm 0.047 \pm 0.052. \cr}}
Taken in conjunction with the isospin relation
\eqn\isospinreln{
R_\pi^+  = {\Gamma({D^*}^+ \to D^0 \pi^+) \over \Gamma({D^*}^+ \to
  D^+ \pi^0)} = 2.21 \pm 0.07, }
these data yield the following branching fractions:
\foot{These very recent CLEO values differ significantly from Particle Data
Group world averages \PDB.}
\eqna\branchingratios
$$ \eqalignno{
{D^*}^+ &\to D^0 \pi^+ \qquad\qquad 68.1 \pm 1.0 \pm 1.3 \% &
  \branchingratios a \cr
{D^*}^+ &\to D^+ \pi^0 \qquad\qquad 30.8 \pm 0.4 \pm 0.8 \% &
  \branchingratios b\cr
{D^*}^+ &\to D^+ \gamma \quad\qquad\qquad 1.1 \pm 1.4 \pm 1.6 \% &
  \branchingratios c\cr
&&\cr
{D^*}^0 &\to D^0 \pi^0 \qquad\qquad 63.6 \pm 2.3 \pm 3.3 \% &
  \branchingratios d\cr
{D^*}^0 &\to D^0 \gamma \space\qquad\qquad 36.4 \pm 2.3 \pm 3.3 \%. &
  \branchingratios e \cr } $$

      Using the branching fraction ratios for the two independent $D^*$
charge modes in \branchingfractionratios, we can deduce the parameters
$\cH/\LX$ and $g_1^2$ that enter into the heavy meson electromagnetic and
strong decay rates respectively.  To extract these unknown couplings from the
$D^*$ data and to predict the $B^*$ widths, we must specify
numerical values for the charm and bottom mass parameters $\mc$ and $\mb$.
Since these quark masses are sources of large theoretical uncertainty,
we perform the fit twice.  First we assume $(\mc,\mb) =
(1500 \MeV,4500 \MeV)$, and then we take
$(\mc,\mb) = (1700 \MeV, 5000 \MeV)$.
Reasonable estimates for the heavy quark masses are covered by the range
between these two sets of input values.

      From the charm vector meson ratios, we find two equations for the
two unknowns:
\eqn\couplings{\eqalign{
& g_1^{-2} \Biggl[ {1 \over 2 \mc} \aEM(\mc)^{1/2} - {\cH \over \LX}
  \aEM(\LX)^{1/2} \Biggr]^2 = \cr
&\qquad\qquad {9 \over 128 \pi f^2}
  {\Bigl[ \bigl(M^2_{{D^*}^+}-(M_{D^+}+m_{\pi^0})^2 \bigr)
          \bigl(M^2_{{D^*}^+}-(M_{D^+}-m_{\pi^0})^2 \bigr) \Bigr]^{3/2}
   \over  \bigl(M^2_{{D^*}^+}-M^2_{D^+} \bigr)^3} \> R_\gamma^+ \cr
& g_1^{-2} \Biggl[ {1 \over 4\mc} \aEM(\mc)^{1/2} +
  {\cH \over \LX} \aEM(\LX)^{1/2} \Biggr]^2 = \cr
&\qquad\qquad {9 \over 512 \pi f^2}
  {\Bigl[ \bigl(M^2_{{D^*}^0}-(M_{D^0}+m_{\pi^0})^2 \bigr)
          \bigl(M^2_{{D^*}^0}-(M_{D^0}-m_{\pi^0})^2 \bigr) \Bigr]^{3/2}
   \over  \bigl(M^2_{{D^*}^0}-M^2_{D^0} \bigr)^3} \> R_\gamma^0 . \cr}}
Following the suggestion of naive dimensional analysis, we choose the
roots of these quadratic equations that yield values for
$g_1^2$ of order unity.  The results of the parameter fit are then listed
as functions of the charm quark mass in Table~1:
$$ \vbox{\offinterlineskip
\def\tablerule{\noalign{\hrule}}
\hrule
\halign {\vrule#& \strut#&
\ \hfil#\hfil& \vrule#&
\ \hfil#\hfil& \vrule#&
\ \hfil#\hfil\ & \vrule# \cr
\tablerule
height10pt && \omit && \omit && \omit &\cr
&& \quad Coupling \quad && \quad $\mc = 1500 \MeV$ \quad
&& \quad $\mc = 1700 \MeV$ \quad & \cr
height10pt && \omit && \omit && \omit &\cr
\tablerule
height10pt && \omit && \omit && \omit &\cr
&& $\cH/\LX$ && $(-0.68 \pm 0.50) / (1000 \MeV)$ &&
  $(-0.60 \pm 0.44) / (1000 \MeV)$ &\cr
height10pt && \omit && \omit && \omit &\cr
&& $g_1^2$ && $0.43 \pm 0.61 $ && $0.34 \pm 0.48 $ &\cr
height10pt && \omit && \omit && \omit &\cr
\tablerule}} $$
\centerline{Table 1}
\bigskip\noindent
These results only weakly depend upon $\LX$ through the
logarithmic running of the fine structure constant.  Therefore, a very
precise numerical value for the chiral symmetry breaking scale need not be
specified.  However, if one reasonably assumes $\LX \approx 1000 \MeV$, then
the value for $\cH$ turns out to be of order one and is consistent with
our earlier expectations.   We also note for comparison that the
nonrelativistic quark model estimate for the squared Goldstone axial vector
parameter is $0.7 \, \ltap \, g_1^2 \, \ltap \, 1.0$ \refs{\Yan,\IsgurWise}.
The HHCPT central value for this coupling is therefore of the same order of
magnitude but smaller than the quark model number.  The large error bars on
$g_1^2$ reflect the $200 \%$ uncertainty in the measurement
\branchingratios{c} of the ${D^*}^+ \to D^+ \gamma$ branching fraction.
Improvements in the experimental value will yield more precise
estimates for this basic chiral Lagrangian parameter.

      Having found $\cH/\LX$ and $g_1^2$, we can now obtain model
independent predictions for the total and partial widths of all $D^*$ and
$B^*$ vector mesons.  Our predictions are summarized in Table~2:
\vfill\eject
$$ \vbox{\offinterlineskip
\def\tablerule{\noalign{\hrule}}
\hrule
\halign {\vrule#& \strut#&
\ \hfil#\hfil& \vrule#&
\ \hfil#\hfil& \vrule#&
\ \hfil#\hfil\ & \vrule# \cr
\tablerule
height10pt && \omit && \omit && \omit &\cr
&& \quad Width $(\MeV)$  \quad && \quad $\mc = 1500 \MeV$ \quad
&& \quad $\mc = 1700 \MeV$ \quad & \cr
&& && \quad $\mb = 4500 \MeV$ \quad && \quad $\mb = 5000 \MeV$ \quad & \cr
height10pt && \omit && \omit && \omit &\cr
\tablerule
height10pt && \omit && \omit && \omit &\cr
&& $\Gamma({D^*}^+)$ && $(12.44 \pm 12.27) \times 10^{-2}$ &&
  $(9.70 \pm 9.56) \times 10^{-2}$ &\cr
height10pt && \omit && \omit && \omit &\cr
&& $\Gamma({D^*}^+ \to D^+ \pi^0)$ && $(3.56 \pm 5.06) \times 10^{-2}$ &&
   $(2.77 \pm 3.94) \times 10^{-2}$ &\cr
height10pt && \omit && \omit && \omit &\cr
&& $\Gamma({D^*}^+ \to D^0 \pi^+)$ && $(7.83 \pm 11.13) \times 10^{-2}$ &&
   $(6.10 \pm 8.68) \times 10^{-2}$ &\cr
height10pt && \omit && \omit && \omit &\cr
&& $\Gamma({D^*}^+ \to D^+ \gamma)$ && $(1.06 \pm 1.05) \times 10^{-2}$ &&
   $(0.83 \pm 0.81) \times 10^{-2}$ &\cr
height10pt && \omit && \omit && \omit &\cr
\tablerule
height10pt && \omit && \omit && \omit &\cr
&& $\Gamma({D^*}^0)$ && $(6.49 \pm 7.94) \times 10^{-2}$ &&
   $(5.06 \pm 6.19) \times 10^{-2}$ & \cr
height10pt && \omit && \omit && \omit &\cr
&& $\Gamma({D^*}^0 \to D^0 \pi^0)$ && $(5.36 \pm 7.63) \times 10^{-2}$ &&
   $(4.18 \pm 5.94) \times 10^{-2}$ &\cr
height10pt && \omit && \omit && \omit &\cr
&& $\Gamma({D^*}^0 \to D^0 \gamma)$ && $(1.13 \pm 2.20) \times 10^{-2}$ &&
   $(0.88 \pm 1.71) \times 10^{-2}$ &\cr
height10pt && \omit && \omit && \omit &\cr
\tablerule
height10pt && \omit && \omit && \omit &\cr
&& $\Gamma({{{\bar{B}}^*}^+}) = \Gamma({{{\bar{B}}^*}^+} \to
  {\bar B}^+ \gamma)$ && $(8.46 \pm 11.94) \times 10^{-4}$ &&
  $(6.60 \pm 9.31) \times 10^{-4}$ & \cr
height10pt && \omit && \omit && \omit &\cr
\tablerule
height10pt && \omit && \omit && \omit &\cr
&& $\Gamma({{{\bar{B}}^*}^0}) = \Gamma({{{\bar{B}}^*}^0} \to
  {\bar B}^0 \gamma)$ && $(1.63 \pm 2.61) \times 10^{-4}$ &&
  $(1.27 \pm 2.03) \times 10^{-4}$ &\cr
height10pt && \omit && \omit && \omit &\cr
\tablerule}} $$
\centerline{Table 2}
\bigskip\noindent
Current upper bounds on $D^*$ widths are about an order of magnitude
greater than the central values quoted here, while no $B^*$ decay
information is yet available.  Comparison of these theoretical results
with experimental data must therefore be left for the future.

      To conclude, we comment upon several possible extensions of this
work. In the meson sector, a number of refinements of our
leading order analysis should be pursued.  Perturbative QCD corrections,
subleading $O(1/\mQ)$ and $SU(3)_{\L+\R}$ breaking effects, and calculable
nonanalytic terms from Goldstone boson loop diagrams may all be systematically
incorporated into the HHCPT framework to yield improved values for the
meson parameters and decay rates.  $D_s^*$ and $B_s^*$ decays can also be
worked out and studied in a straightforward fashion. For the sextet baryons,
the present absence of branching ratio data precludes our determining the
baryon couplings $(\cS/\LX, g_2^2)$ and $(\cST/\LX, g_3^2)$ as well as the
widths of the spin-$\threehalves$ states in precisely the same manner as
their meson analogues.  Nonetheless, such baryon data will eventually become
available in the future.  So the enhancements mentioned above for the
mesons ought to be carried out for the baryons as well.  Finally, the
scope of HHCPT can be broadened to include higher resonances such as the
$D_1$ and $D_2^*$ states \FalkLuke.  Electromagnetic interactions for these
meson and baryon excitations may be incorporated into the theory along the
same lines as those for the heavy hadron $H'$, $S$ and $T$ ground states.

\bigskip
\centerline{\bf Acknowledgements}
\bigskip

      We thank Glenn Boyd and Mark Wise for several discussions and for
communicating results prior to publication.  These authors and their
collaborators have independently derived many of the findings reported here
\Amundson.  We are also grateful to Steve Schaffner, Mat Selen and Hitoshi
Yamamoto for providing access to CLEO data.  Finally, PC thanks the theory
group at Fermilab where part of the work on this letter was performed for
their warm hospitality.  This work was supported in part by the National
Science Foundation under contract PHY-87-14654 and by the Texas National
Research Commission under Grant \# RGFY9206.

\listrefs
\bye